\documentclass[prl,showpacs,twocolumn]{revtex4}

\usepackage{graphics}
\usepackage{amsmath}
\usepackage{slashed}
\usepackage{feynmp}

\newcommand{\Tr}{\mathop{\rm Tr}\nolimits}% trace operator

\begin{document}
\title{A model that underlies the Standard model}
\author{Ji\v r\'{\i} Ho\v sek}
\email{hosek@ujf.cas.cz} \affiliation{Department of Theoretical
Physics, Nuclear Physics Institute, Czech Academy of Sciences, 25068
\v Re\v z (Prague), Czech Republic}

\begin{abstract}
We assign the chiral fermion fields of the Standard model to
triplets of flavor (family, generation, horizontal) $SU(3)_f$
symmetry, for anomaly freedom add one triplet of sterile
right-handed neutrino fields, and gauge that symmetry. First we
demonstrate that the resulting quantum flavor $SU(3)_f$ dynamics
completely spontaneously self-breaks: Both the Majorana masses of
sterile neutrinos and the masses of all eight flavor gluons come out
proportional to the $SU(3)_f$ scale $\Lambda$. Mixing of sterile
neutrinos yields new CP-violating phases needed for understanding
the baryon asymmetry of the Universe. Second, the $SU(3)_f$ dynamics
with weak hypercharge radiative corrections spontaneously generates
the lepton and quark masses exponentially suppressed with respect to
$\Lambda$. Three active neutrinos come out as Majorana particles
extremely light by seesaw. The Goldstone theorem implies: (i) The
electroweak bosons $W$ and $Z$ acquire masses. (ii) There are three
axions, decent candidates for dark matter. Invisibility of the
Weinberg-Wilczek axion $a$ with mass $m_a \sim \rm
m^2_{\pi}/\Lambda$ restricts the scale $\Lambda$ from $\Lambda \sim
\rm 10^{10} GeV$ upwards. Third, the composite 'would-be'
Nambu-Goldstone (NG) bosons of all spontaneously broken gauge
symmetries have their {\it genuine composite} massive partners: (i)
One $0^{+}$ flavorless Higgs-like particle $h$ accompanying three
electroweak 'would-be' NG bosons. (ii) Two $0^{+}$ flavored
Higgs-like particles $h_3$ and $h_8$ accompanying six flavored
electroweak 'would-be' NG bosons. (iii) Three superheavy spin-zero
sterile-neutrino-composites $\chi_i$ accompanying eight flavored
sterile-neutrino-composite 'would-be' NG bosons. We identify
$\chi_i$ with inflatons.
\end{abstract}


\maketitle

\section{I. Introduction and Summary}
Standard model (SM) as a quantum field theory is firmly based on two
general principles: The gauge principle, and the principle of
spontaneous symmetry breaking. Principles are, however, more general
than their particular realizations \cite{lee}. The SM realization of
the gauge principle, defining the gauge particles of the underlying
symmetries and fixing the form of their interactions with matter
fields and with themselves is in full accord with data. The
electroweak Higgs realization of the principle of spontaneous
symmetry breaking giving particles softly their masses is less
certain: Glorious CERN LHC discovery \cite{lhc} of the spinless
$0^{+}$ $125 \phantom{i} \rm GeV$ boson with properties similar to
the SM Higgs certainly does support the Higgs realization,
technically all the way up to the Planck scale. It is, however, far
from complete: First, it does not provide enough CP violation needed
for the observed baryon asymmetry of the Universe. Second, it leaves
neutrinos massless. Third, it does not offer any candidates for
particles of dark matter. Fourth, it does not describe inflation of
the early Universe. Fifth, if the CERN Higgs were indeed the Higgs
boson of the Standard model the masses of quarks and charged leptons
would stay {\it theoretically arbitrary} for ever.

This 'environmental' interpretation of the SM fermion mass spectrum
is in sharp contrast with our understanding of the energy spectra of
other quantum systems: Oscillators, nuclei, atoms and molecules have
their spectra {\it calculable}. In the same vein, the spectrum of
hadron masses is calculable in QCD in the chiral limit solely in
terms of the QCD scale. That the laws of QCD at low momenta are
known at present only to computers is another issue.

Here we suggest to replace the essentially classical Higgs sector of
the SM with its 'twenty-some' parameters \cite{lee} by a new
genuinely quantum non-Abelian dynamics. It is defined by gauging the
flavor (family, generation, horizontal) $SU(3)_f$ triplet index of
three chiral SM lepton ($l_{fL}, e_{fR}$) and quark ($q_{fL},
u_{fR}, d_{fR}$) families of the $SU(2)_L \times U(1)_Y$ gauge
invariant SM. This amounts to introduction of the octet of gauge
flavor gluons $C_a^{\mu}$, and for anomaly freedom to addition of
one triplet of sterile right-handed neutrino fields $\nu_{fR}$. The
resulting anomaly free, asymptotically free gauge $SU(3)_f$ quantum
flavor dynamics is characterized by one parameter. It is either the
dimensionless gauge coupling constant $h$ or, due to the dimensional
transmutation, the theoretically arbitrary scale $\Lambda$. Its
Lagrangian is
\begin{eqnarray*}
{\cal L}_f=-\frac{1}{4}F_{a\mu\nu}F_a^{\mu\nu}+ \ \bar q_L i\slashed
D q_L + \bar u_R i\slashed D u_R + \bar d_R i\slashed D d_R\\ + \bar
l_L i\slashed D l_L + \bar e_R i\slashed D e_R + \bar \nu_{R}
i\slashed D \nu_{R}
\end{eqnarray*}
\noindent Treated in isolation ${\cal L}_f$ should be appended by
the $SU(3)_f$ invariant hard Dirac fermion mass term
\begin{equation*}
{\cal L}_{mass}=-\sum_f (\bar f_R m_f f_L + h.c.)
\end{equation*}
common to all three fermions $f=u, d, e, \nu$ of a given electric
charge. Such terms are, however, strictly prohibited by the gauge
electroweak chiral $SU(2)_L \times U(1)_Y$ symmetry tacitly always
present in the game.

The form of ${\cal L}_f$ is identical with the form of the QCD
Lagrangian in the chiral limit. This is highly suspicious. 'In QCD
we trust', and it is a firm experimental fact that the flavor
symmetry is not confining but badly broken.

Closer inspection reveals that the presence of the kinetic term
$\bar \nu_{R} i\slashed D \nu_{R}$ of the {\it electrically neutral}
right-handed neutrinos makes the cardinal difference from QCD: The
flavor gluon interaction likes to generate at the strong coupling
dynamically the Majorana mass term
\begin{equation}
{\cal L}_{Majorana}=-\tfrac{1}{2}(\bar \nu_{R}M_{R}(\nu_{R})^{{\cal
C}}+ h.c.)\label{majorana}
\end{equation}
where $M_{iR}$ are three different Majorana masses of order
$\Lambda$. There is no way how ${\cal L}_{Majorana}$ can be
$SU(3)_f$ invariant, so it is strictly prohibited at the Lagrangian
level as a hard mass term. The Goldstone theorem applies, and the
resulting composite 'would-be' NG bosons give rise to different
masses of all eight flavor gluons $C$ proportional to $M_{iR}$.
Hence the gauge $SU(3)_f$ symmetry of ${\cal L}_f$ gets dynamically
completely self-broken. Because $(\nu_{R})^{{\cal C}}$ is a
left-handed field transforming as an {\it antitriplet} of $SU(3)_f$,
the new dynamics is not QCD-like, but it is effectively chiral.

We will argue that the flavor gluon dynamics generates also the
masses of the electroweakly interacting fermions and, due to the
Goldstone theorem again the electroweak gauge symmetry is
spontaneously broken down to $U(1)_{\rm em}$. The underlying
'would-be' NG bosons give rise to masses of the $W$ and $Z$ bosons.

Since it is not known at present how to put a chiral gauge theory on
the lattice \cite{kaplan} the present attempt at computing the
fermion mass spectrum should only be considered as a heuristic
prototype computation. We believe nevertheless that it has all
necessary attributes expected by common sense.

For analyzing the consequences of the fermion mass calculation we
will borrow the strategy successfully applied in QCD in the $SU(2)_L
\times SU(2)_R$ chiral limit: The detailed properties of the QCD
ground state responsible for the confinement of its colored
constituents are not known. {\it If}, however, we {\it assume} that
the essentially unknown vacuum breaks the global chiral symmetry
spontaneously, the {\it composite massless colorless NG pions}
described in the effective chiral perturbation theory \cite{hl} are
obliged to exist by the existence theorem of Goldstone. To the best
of our knowledge there is no existence theorem for other composite
colorless (massive) hadrons. Because hadrons are phenomenologically
so important many models of their formation were invented over the
years. Grasping different expected properties of the confining
vacuum they correspondingly differ: from the constituent quark model
to the bag models of different types.

We will argue analogously: It is not known how the nonconfining
vacuum of strongly coupled flavor dynamics acts in the infrared.
{\it If}, however, we {\it assume} that it generates appropriate
chirality changing fermion proper self energies dynamically, the
existence theorem of Goldstone implies definite firm predictions.
There must be the whole spectrum of the true, 'would-be' and pseudo
NG bosons of underlying global anomaly-free, local anomaly free, and
global anomalous Abelian chiral symmetries with properties fixed by
symmetry. The basic assumption thus implies:

(1) There is one true composite NG boson of a spontaneously broken
global anomaly-free Abelian symmetry of the model. It is rather
remarkable that the prediction of a massless spinless particle is
not in flagrant conflict with data. The point is that it does not
imply a new long-range force \cite{gelmini}. Moreover, it is always
possible, if the experimental data demand, to gauge that symmetry.
In such a case the 'would-be' NG boson disappears: We would cope
with a new $Z'$ characterized by a new gauge coupling $g''$ with
mass proportional to the masses of all fermions to which $Z'$
couples.

(2) There must be the massive gauge bosons of gauge chiral
symmetries $SU(2)_L \times U(1)_Y$ and $SU(3)_f$. The 'would-be' NG
bosons disappear from the physical spectrum and become the
longitudinal polarization states of massive $W,Z$ and of the flavor
gluons $C$ \cite{jj}. In the case of the electroweak $SU(2)_L \times
U(1)_Y$ there is a relation between Dirac masses of the
electroweakly interacting fermions and the masses of $W$ and $Z$. In
the case of gauge flavor dynamics the phenomenological viability
requires that the {\it flavor changing} flavor gluons get very heavy
masses. This comes out very naturally: The flavor gluon masses are
the inevitable consequence of huge masses of sterile right-handed
Majorana neutrinos.

(3) There must be three composite pseudo NG bosons of spontaneously
broken global Abelian chiral anomalous symmetries. They acquire
masses by instantons of three non-Abelian gauge forces present in
the model. One of them we identify with the Weinberg-Wilczek axion
\cite{ww}, most welcome particle postulated originally to solve the
strong CP problem. Some like it also as a hot candidate for cold
dark matter \cite{sikivie}. Other particle is an ultra-light
electroweak axion also postulated previously by Anselm and Uraltsev
\cite{au}. The third one is a new axion of quantum flavor dynamics.
We fix its mass wishfully in the $\rm keV$ range, and offer it as a
candidate for the explanation of several astrophysical puzzles
\cite{ringwald}.

Besides the collective excitations guaranteed by the Goldstone
theorem it is natural to expect other massive composites of a
strongly coupled dynamics. After the discovery of the Higgs boson
this possibility became necessity. To the best of our knowledge
there is no existence theorem for such particles i.e., the spectrum
of composite non-NG excitations is model-dependent. Guided by the
canonical Higgs model we look for the {\it composite scalar fields
with the SM Higgs field quantum numbers which can condense}. We
emphasize that the Standard model with its canonical Higgs sector is
phenomenologically so successful that the existence of such a
composite operator should be the {\it necessary condition} for the
viability of the model.

The composite Higgs $h$ is a massive partner of three composite
'would-be' NG bosons which follow from the spontaneous electroweak
symmetry breakdown by dynamically generated lepton and quark masses.
They are identified in the composite scalar field by other means (by
the Ward-Takahashi (WT) identity). Its Yukawa interaction extracted
from the 'partnership' is identical with the SM one. Its
interactions with the electroweak gauge bosons $W, Z, A$ are {\it
all} the effective ones and uniquely come out from the ultraviolet
(UV)-finite fermion loops.

Following the same reasoning we find analogous operator also in the
sterile neutrino sector. It is a complex composite flavor sextet of
$SU(3)_f$. Besides the eight composite 'would-be' NG bosons which
follow from the spontaneous breakdown of $SU(3)_f$ by dynamically
generated Majorana masses of sterile right-handed neutrinos it
contains in this case three Higgs-like massive composite excitations
$\chi_i(x)$. Twelveth component is the composite pseudo NG boson
discussed in Sect.V.

To follow the same reasoning consistently we are enforced to
consider also the composite multicomponent Higgs field associated
with spontaneous breakdown of $SU(3)_f$ by masses of the
electroweakly interacting leptons and quarks. It turns out that it
is a composite octet of $SU(3)_f$. Because the Higgs octet
condensate breaks in general the $SU(3)$ down to unbroken $U(1)
\times U(1)$, the two composite Higgs-like particles $h_3$ and $h_8$
should remain in the physical spectrum of the model. Because the
lepton and quark masses are tiny in comparison with huge Majorana
masses of sterile neutrinos, the contributions of the underlying
'would-be' NG bosons to the masses of flavor gluons are neglected.

In contrast to the canonical SM Higgs boson the effective
interactions of the composite Higgs-like particles are calculable.
This conclusion is supported by the work \cite{kuti} arguing in
favor of equivalence of the weakly coupled model with the elementary
scalar Higgs field and the strongly coupled renown BHL model
\cite{bhl}. It will become obvious in the following that our model
comes close to the BHL one in a very crude low-momentum
approximation.

Finally, it is of course mandatory to demonstrate that our basic
assumption of the dynamical fermion mass generation is warranted. We
support this assumption by finding explicitly, in separable
approximation, the matrix chirality-changing lepton and quark proper
self energies $\Sigma(p^2)$ as the nonperturbative strong-coupling
solution of the Schwinger-Dyson (SD) equation. Separable Ansatz for
the kernel represents a particular model of the momentum-dependent
coupling $\bar h_{ab}^2(q^2)$ of the gauge $SU(3)_f$ in the
infrared, and enables to derive the exponentially sensitive formula
for lepton and quark masses
\begin{equation*}
m_i=\Lambda \phantom{b} \rm exp (-1/4\alpha_i)
\end{equation*}
in terms of a handful of the effective low-momentum constants
$\alpha_i$. It is gratifying that at the same time the identical
approximation naturally yields the huge Majorana masses of all three
sterile right-handed neutrinos.

The idea of gauging the flavor or family or generation or horizontal
symmetry is so natural that it could hardly be new
\cite{gaugedfamilysymmetry}: It ties together a number of
troublesome global non-Abelian symmetries which necessarily emerge
once the standard Higgs sector with its general Yukawa couplings is
switched off. At the same time it is obvious that: (i) such a
symmetry is badly broken, and (ii) its breakdown cannot be explicit.
In all papers on this subject known to us this gauge flavor symmetry
is spontaneously broken by an extended Higgs sector of elementary
condensing scalar fields. We argue that no elementary scalars are
necessary i.e., that the gauge flavor $SU(3)_f$ dynamics, strongly
coupled in the infrared, {\it due to its sterile neutrino sector},
completely self-breaks. It should be noted that the idea of the
dynamical breakdown of $SU(3)_f \times SU(2)_L \times U(1)_Y$
symmetry is mentioned by Yanagida in \cite{gaugedfamilysymmetry}.

The paper is structured as follows. In Sect.II we demonstrate that
the sole sector of sterile right-handed neutrinos yields the
complete self-breaking of the gauge chiral $SU(3)_f$: The flavor
gluon exchanges at low momenta between the right-handed sterile
neutrino fields and their charge-conjugate left-handed counter parts
generate three different Majorana masses $M_R$. They follow from the
chiral symmetry breaking parts $\Sigma(p^2)$ found in a separable
approximation as the low-momentum solutions of the Schwinger-Dyson
equation for the full matrix neutrino propagator. All flavor gluons
then necessarily acquire masses by absorbing the composite
'would-be' NG bosons as their longitudinal polarization states. It
is gratifying that the formalism is convincingly supported by the
description of spontaneous Majorana mass generation and of the
consequent complete breakdown of gauge $SU(3)$ using the elementary
scalar field. All what is needed is one Higgs field in the complex
symmetric sextet representation of $SU(3)$ \cite{bhs}.

Sect.III is devoted to the aspects of electroweak symmetry breaking
in the present model: First, we solve the Schwinger-Dyson equations
for the Dirac chirality-changing proper self energies $\Sigma(p^2)$
of the SM fermions in separable approximation. Then we demonstrate
that they result in a wide and wild spectrum of lepton and quark
masses. Second, we compute the $W$ and $Z$ boson masses in terms of
the fermion self-energies $\Sigma(p^2)$.

In Sect.IV we reveal the composite Higgs boson as a massive partner
of the 'would-be' composite electroweak NG bosons, and indicate the
derivation of its effective interactions with fermions, and with the
electroweak gauge bosons. Similar argumentation leads to the
expectation of the existence of new Higgs-like bosons $h_{3}$ and
$h_{8}$.

In Sect.V we briefly analyze the basic properties of three composite
axions, the pseudo NG bosons created by global anomalous Abelian
currents. Axions acquire their masses by the non-perturbative
effects of three different non-Abelian gauge interactions present in
the model.

In Sect.VI we collect several specific phenomenological consequences
of the model, and Sect.VII contains our conclusions and an outlook.

\section{II. Fertile sterile neutrinos}

In perturbation theory the Lagrangian of the sterile neutrino sector

\begin{eqnarray}
{\cal L}_{\nu_R}=\bar \nu_{R}i\slashed \partial \nu_{R} + h\bar
\nu_{R}\gamma_{\mu}\tfrac{1}{2}\lambda_a\nu_{R}C^{\mu}_a
-\frac{1}{4}F_{a\mu\nu}F_a^{\mu\nu} \label{nuR}
\end{eqnarray}

\noindent describes the triplet of massless right-handed neutrinos
$\nu_{R}$ interacting with the octet of massless spin-1 flavor
gluons $C_a^{\mu}$ in accordance with exact $SU(3)_f$ gauge
invariance (anomaly freedom is ignored for the moment). The hard
Majorana mass term $\bar \nu_R M_R (\nu_R)^{{\cal C}}$ is not the
$SU(3)_f$ singlet ($\bar 3 \times \bar 3 = 3_a + \bar 6_s$) and is
therefore strictly prohibited by the $SU(3)_f$ symmetry. (The
subscripts $a,s$ abbreviate the antisymmetric and symmetric
representations, respectively.) The Lagrangian (\ref{nuR}) obviously
obeys also the global $U(1)_s$ 'sterility' symmetry.

We bring reasonable nonperturbative arguments that the dynamics
defined by (\ref{nuR}) in its strong coupling low-momentum regime is
not confining, but it yields the complete spontaneous self-breaking:
First, (\ref{nuR}) describes three Majorana neutrinos with different
masses $M_R$ of order $\Lambda$. Second, these dynamically generated
different masses break the $U(3)_f$ symmetry spontaneously and
completely. (i) Since the chiral $SU(3)_f$ is the gauge symmetry,
its spontaneous breakdown generates eight composite 'would-be' NG
bosons. They give rise to masses $M$ of all eight flavor gluons
proportional to $M_R$. (ii) The composite NG boson of the global
Abelian chiral symmetry remains in the spectrum. We will deal with
it in detail in Sect.V. taking into account the axial anomaly.
Third, as a remnant of the just described dynamical Higgs mechanism
there should be three massive composite Higgs-like particles with
calculable effective interactions with flavor gluons, with massive
Majorana neutrinos and with themselves. Fourth, because below
$\Lambda$ the dynamics is strongly coupled and nonconfining, it is
natural to expect that it generates other massive composite
excitations. In particular, the flavored ones should contribute to
the flavor gluon polarization tensor. This last point is mentioned
here for its importance but its elaboration is beyond the scope of
this paper. We use this expectation for justification of the Ansatz
for the momentum-dependent sliding coupling $\bar h_{ab}^2(q^2)$ in
(\ref{Sigma}).

The impressionistic picture painted above is not unexpected: The
massless fields can excite massive particles. This comes about by
finding nonperturbatively the nontrivial poles of their full
propagators. For fermions for which the masslessness is protected by
the chiral symmetry this amounts to finding the chiral symmetry
breaking self energies in their full propagators \cite{njl}. For
gauge bosons for which the masslessness is protected by the gauge
symmetry this amounts, according to Schwinger \cite{schwinger}, to
finding the residue at the massless pole of the gauge field
polarization tensor. Moreover, it is a common knowledge that there
exists the convincing (say phenomenological) realization of this
picture, using the standard Higgs mechanism. Close relation between
the strong-coupling microscopic and the weak-coupling
phenomenological descriptions allows us to predict the existence of
three massive composite Higgs-like particles which in the
microscopic dynamics is difficult to identify.

We employ below the non-perturbative self-consistency reasoning
underlying the concept of spontaneous symmetry breaking pioneered by
Yoichiro Nambu: First we {\it assume} that three different $SU(3)_f$
symmetry breaking Majorana masses of $\nu_R$ are dynamically
generated. This implies the existence of eight 'would-be' NG
composite excitations which in turn give rise to different masses of
all eight flavor gluons $C$. Massive flavor gluons then imply the
symmetry-breaking kernel in the Schwinger-Dyson equation for the
fermion masses. Finally, using this form of the SD equation we {\it
find} explicitly the assumed Majorana masses.

This genuinely quantum complete dynamical breakdown of the
strong-coupling gauge $SU(3)_f$ symmetry is considerably more
complicated than the standard, essentially classical, Higgs
realization: Using the elementary Higgs field $\Phi$ in the complex
symmetric sextet representation of $SU(3)_f$ we construct the
manifestly gauge $SU(3)_f$ invariant Lagrangian using all three
algebraically independent invariants which can be constructed from
$\Phi$. With appropriate choice of parameters in the Lagrangian the
classical Hamiltonian has the minimum corresponding to the vacuum
expectation value of $\Phi$ with three different diagonal nonzero
positive entries. It is then a manual work to demonstrate that such
a condensate generates three different Majorana neutrino masses and
eight different masses of flavor gluons. As a bonus there are {\it
necessarily} three Higgs-like $0^{+}$ scalars with different masses
and prescribed interactions.

\subsection{1. Dynamical Majorana neutrino mass generation}

To see the possibility of the dynamical Majorana neutrino mass
generation we rewrite the neutrino-flavor gluon interaction in
(\ref{nuR}) identically as
\begin{eqnarray}
{\cal L}_{\rm int}=\tfrac{1}{2}h\{\bar
\nu_{R}\gamma_{\mu}\tfrac{1}{2}\lambda_a\nu_{R}+ \overline
{(\nu_{R})^{{\cal
C}}}\gamma_{\mu}[-\tfrac{1}{2}\lambda_a^T](\nu_{R})^{{\cal
C}}\}C^{\mu}_a \label{n}
\end{eqnarray}
where we have introduced the charge conjugate neutrino field
$(\nu_R)^{{\cal C}}= C (\bar \nu_R)^T$. It is important to realize
that it is a {\it left-handed field}, $(\nu_R)^{{\cal C}}=(\nu^{\cal
C})_L$, transforming as the {\it antitriplet} of $SU(3)_f$:
$T_a(L)=-\tfrac{1}{2}\lambda_a^{T}$. These ingredients are simply
necessary: Any fermion mass term is a bridge between the left- and
the right-handed fermion fields. In the present case these are not
independent fields, but are related by charge conjugation. Moreover,
because the left-handed neutrino field transforms as an antitriplet
and the right-handed one as a triplet, the $SU(3)_f$ dynamics is not
vector-like. As a result the Majorana mass matrix (or, more
generally the Majorana chiral symmetry breaking self-energy), if
dynamically generated, must be a general $3 \times 3$ matrix {\it
symmetric} by Pauli principle.

We consider in this paper the full matrix fermion propagator $S(p)$
(with unimportant fermion wave function renormalization set to one)
in the form devised by Petr Bene\v s \cite{benes1}
\begin{eqnarray}
S^{-1}(p)=\slashed p - \hat \Sigma(p^2) \label{s}
\end{eqnarray}
where $\hat \Sigma=\Sigma P_L+\Sigma^{+}P_R$ and
$P_{L,R}=\tfrac{1}{2}(1\mp\gamma_5)$.

For sterile neutrinos $S^{-1}(p)$ corresponds to the effective
bilinear Lagrangian \cite{benes2}
\begin{eqnarray}
{\cal L}_{eff}^{(2)}&=&\tfrac{1}{2} \bar \nu_R  \slashed p \nu_R
+\tfrac{1}{2}\overline{(\nu_{R})^{{\cal C}}} \slashed
p(\nu_{R})^{{\cal C}}-\tfrac{1}{2}[( \bar \nu_R \Sigma
{(\nu_{R})^{{\cal C}}}+ h.c.)]\nonumber\\
&=&\tfrac{1}{2}\bar n \slashed p n -\tfrac{1}{2}\bar n \hat \Sigma n
\label{leff}
\end{eqnarray}
We have introduced the Majorana neutrino field $n=\nu_R
+(\nu_{R})^{{\cal C}}$ having the obvious property $n^{{\cal C}}=n$.

It is utmost important that Bene\v s's form of the matrix fermion
propagator $S^{-1}$ enables to deal explicitly with the
'denominator' of the matrix $S(p)$ and to analyze technically the
phenomenologically important program of fermion mixing:
\begin{equation*}
S(p)=(\slashed p+\Sigma^{+})(p^2-\Sigma\Sigma^{+})^{-1}P_L+(\slashed
p+\Sigma)(p^2-\Sigma^{+}\Sigma)^{-1}P_R
\end{equation*}
\vspace{1mm}

We say that the massless neutrino fields $\nu_R$ create massive
Majorana neutrinos if nonzero chiral symmetry changing $\hat
\Sigma(p^2)$ is found as an ultraviolet (UV) finite solution of the
Schwinger-Dyson equation \cite{pagels}
\begin{widetext}
\begin{eqnarray}
\hat \Sigma(p)=3\int \frac{d^4k}{(2\pi)^4}\frac{\bar
h^2_{ab}((p-k)^2)}{(p-k)^2}T_a(R)\{\Sigma(k)[k^2+\Sigma^{+}(k)\Sigma(k)]^{-1}P_L+\Sigma^{+}(k)[k^2+\Sigma(k)\Sigma^{+}(k)]^{-1}P_R\}T_b(L)
\label{Sigma}
\end{eqnarray}
\end{widetext}
We emphasize that the same equation is valid also for the Dirac
fermions. In the latter case the left- and the right-handed fermions
are independent fields, and both transform as triplets. In both
cases the fermion mass spectrum is given by the poles of $S(p)$
i.e., by solving the equation
\begin{equation}
\rm det[p^2-\Sigma(p^2)\Sigma^{+}(p^2)]=0\label{m}
\end{equation}

The sliding coupling $\bar h_{ab}^2(q^2)$ in (\ref{Sigma}) defined
in terms of the flavor gluon polarization tensor contains important
informations about the assumed low-momentum properties of the model.
In particular, it corresponds to the phase in which all flavor
gluons are massive. Despite this, it remains unknown. The point is
that the spectrum of the expected composites carrying flavor, which
by definition below $\Lambda$ contribute to $\bar h_{ab}^2(q^2)$, is
entirely unknown. Finding the fermion spectrum is therefore a
formidable task.

In order to proceed we approximate the problem of finding the
fermion mass spectrum as follows:

(1) In the perturbative weak coupling high-momentum region from
$\Lambda$ to $\infty$ which in technical sense guarantees the UV
finiteness of $\Sigma(p^2)$ \cite{pagels} we set the known
perturbative i.e. small, $\bar h_{ab}^2(q^2)$
\begin{equation*}
\frac{h_{ab}^2(q^2)}{4\pi}=\frac{\delta_{ab}}{(11-\tfrac{n_f}{3})\rm
ln(q^2/\Lambda^2)}
\end{equation*}
equal to zero. Here $n_f=16$ is the number of chiral fermion
triplets. {\it The resulting model is thus not asymptotically, but
strictly free above the scale $\Lambda$.}

We project from (\ref{Sigma}) the equation for $\Sigma$, fix without
loss of generality in the resulting SD equation the external
euclidean momentum as $p=(p,\vec 0)$, integrate over angles and get
\begin{equation}
\Sigma(p)=\int_0^{\Lambda}k^3dk
K_{ab}(p,k)T_a(R)\Sigma(k)[k^2+\Sigma^{+}\Sigma]^{-1}T_b(L)
\label{Sigmasep}
\end{equation}
where the {\it unknown} kernel
\begin{equation}
K_{ab}(p,k)\equiv \frac{3}{4\pi^3}\int_0^{\pi}\frac{\bar
h_{ab}^2(p^2+k^2-2pk \rm \cos \theta)}{p^2+k^2-2pk \rm \cos
\theta}\rm \sin^2 \theta d \rm \theta
\end{equation}
is separately symmetric in momenta and in the flavor octet indices.

(2) Our key approximation is the {\bf separable approximation} for
the kernel $K_{ab}(p,k)$. In the following we analyze explicitly the
Ansatz
\begin{equation}
K_{ab}(p,k)=\frac{3}{4\pi^2} \frac{g_{ab}}{pk}\label{sep}
\end{equation}

The Ansatz substitutes our ignorance of knowing the low-momentum
$\bar h_{ab}^2(q^2)$ {\it and} the low-momentum form of the flavor
gluon propagators (to be found subsequently). Ultimately we should
deal with a system of Schwinger-Dyson equations for several Green
functions, an entirely hopeless task.

Here $g_{ab}$ is a real symmetric matrix of the effective
low-momentum dimensionless coupling constants. They reflect the
complete breakdown of $SU(3)_f$ and are ultimately calculable. We
think they are analogous to the effective low-momentum couplings
\cite{hl} of the chiral perturbation theory of the confining QCD.

Separable approximation has several advantages (approved eventually
a posteriori).

1. The nonlinearity of the integral equation is preserved. We expect
that the non-analyticity of $\Sigma$ upon the couplings $g_{ab}$ is
crucial for generating the huge fermion mass ratios.

2. In separable approximation the homogeneous nonlinear integral
equation (\ref{Sigmasep}) is immediately formally solved:
\begin{equation}
\Sigma(p)=\frac{\Lambda^2}{p}T_a(R)\Gamma_{ab}T_b(L)\equiv\frac{\Lambda^2}{p}\sigma
\label{sol}
\end{equation}
The difficult part is that the numerical matrix $\Gamma$ has to
fulfil the homogeneous nonlinear algebraic self-consistency
condition (gap equation)
\begin{eqnarray}
\Gamma_{ab}&=&g_{ab}\frac{3}{16\pi^2}\int_{0}^{1}dx (T(R)\Gamma
T(L)) \nonumber\\
&& [x + (T(R)\Gamma T(L))^{+}(T(R)\Gamma T(L))]^{-1}\label{Gamma}
\end{eqnarray}

3a. For neutrinos $\Sigma$ describes both the masses of Majorana
neutrinos, and their mixing (including the new CP-violating phases):
The general complex symmetric $3 \times 3$ matrix $\sigma$ can be
put into a positive-definite real diagonal matrix $\gamma$ by a
constant unitary transformation
\begin{equation}
\sigma=U^{+}\gamma U^{*}
\end{equation}
The gap equation becomes
\begin{equation}
\gamma=UT_a(R)U^{+}g_{ab}I(\gamma)U^{*}T_b(L)U^{T} \label{gammaM}
\end{equation}
where
\begin{eqnarray}
I(\gamma)&=&\frac{3}{16\pi^2}\gamma
\int_{0}^{1}\frac{dx}{x+\gamma^2}= \frac{3}{16\pi^2}\gamma \rm
ln\frac{1+\gamma^2}{\gamma^2}\nonumber\\
%&\approx& -\frac{3}{16\pi^2}\gamma \rm ln \gamma^2
\end{eqnarray}

The diagonal entries of the equation (\ref{gammaM}) determine the
sterile neutrino masses, the nondiagonal entries provide relations
for the mixing angles and the new CP-violating phases. These phases
are most welcome as a source of an extra CP violation needed for
understanding of the baryon asymmetry of the Universe
\cite{leptogenesis}.

3b. For Dirac fermions, as in the Standard model, the generally
complex $3 \times 3$ matrix $\sigma$ can be put into a
positive-definite real diagonal matrix $\gamma$ by a constant
bi-unitary transformation:
\begin{equation}
\sigma=U^{+}\gamma V
\end{equation}
The gap equation becomes
\begin{equation}
\gamma=UT_a(R)U^{+}g_{ab}I(\gamma)VT_b(L)V^{+} \label{gammaD}
\end{equation}
The diagonal entries of the equation (\ref{gammaD}) determine the
fermion masses, the nondiagonal entries provide relations for the
CKM mixing angles and the SM CP-violating phase.

\vspace{3mm}

It is natural to demonstrate the existence of the solutions of the
gap equations simultaneously for both the sterile neutrino Majorana
masses and for the Dirac masses of the electroweakly interacting
leptons and quarks. This is done in the Appendix.

\vspace{3mm}

With neutrino mixing neglected and with rather simplifying
assumptions on the effective couplings $g_{ab}$ we obtain in the
Appendix the desirable result
\begin{equation}
\boxed{M_{iR}\sim \Lambda} \label{MR}
\end{equation}
In Sect.V we set the scale $\Lambda$ from the invisibility of the
Weinberg-Wilczek axion $a$.  Within a rather wide window we consider
for definiteness $\Lambda \sim 10^{10} \rm GeV$.

\subsection{2. Self-consistent generation of flavor
gluon masses}

It is self-evident that three different Majorana masses
(self-energies $\Sigma$) break the $U(3)$ symmetry of (\ref{nuR})
completely. Because the appearance of different Majorana masses is
spontaneous, there must be nine NG bosons. Eight of them are the
'would-be', the ninth is the 'pseudo', discussed in Sect.V.

Following \cite{benes2} we reveal the eight 'would-be' NG bosons  as
the massless poles in the Abelian approximation of the
Ward-Takahashi identity for the Green's function
$i\Gamma^{\mu}_a=\langle 0|T[C^{\mu}_a(x)n(y)\bar
n(z)]|0\rangle_{1PI}$ (more precisely for its
one-particle-irreducible part). It is related with the WT identity
for the Green's function $\gamma^{\mu}_a=\langle
0|T[j^{\mu}_a(x)n(y)\bar n(z)]|0\rangle_{1PI}$ associated with
global symmetry generated by the current of Majorana neutrinos $n$:
\begin{eqnarray}
j_a^{\mu}=\bar
\nu_{R}\gamma^{\mu}\tfrac{1}{2}\lambda_a\nu_{R}=\tfrac{1}{2} \bar n
\gamma^{\mu} \tfrac{1}{2}\Lambda_a n \label{current}
\end{eqnarray}
Here
\begin{equation*}
\tfrac{1}{2}\Lambda_a=\tfrac{1}{2}\lambda_a P_R
+\tfrac{1}{2}(-\lambda_a^T) P_L
\end{equation*}
The result is
\begin{eqnarray}
q_{\mu}\Gamma^{\mu}_a(p',p)=\hat
\Sigma(p')\tfrac{1}{2}\Lambda_a-\tfrac{1}{2}\bar \Lambda_a \hat
\Sigma(p)
\end{eqnarray}

The pole term is
\begin{equation}
\Gamma^{\mu}_{a,pole}(p+q,p)=\frac{q^{\mu}}{q^2}[\hat
\Sigma(p+q)\Lambda_a-\bar \Lambda_a\hat \Sigma(p)]\label{pole}
\end{equation}
where
\begin{equation*}
\tfrac{1}{2}\bar \Lambda_a=\gamma_0 \Lambda_a^{+}\gamma_0
\end{equation*}
The term in square brackets describes (with appropriate
normalization) the effective vertices $P_a$ between eight
neutrino-antineutrino composite 'would-be' NG bosons $\pi_a$ and the
corresponding neutrino pair:
\begin{equation}
P_a(p', p)\sim [\hat \Sigma(p')\Lambda_a-\bar \Lambda_a\hat
\Sigma(p)]\label{Pa}
\end{equation}
Not surprisingly we recover the identical structure by computing the
divergence of the current (\ref{current}) using the Dirac equation
with dynamically generated mass following from the bilinear
Lagrangian (\ref{leff}):
\begin{eqnarray*}
(p'-p)_{\mu}\tfrac{1}{2} \bar n(p') \gamma^{\mu}
\tfrac{1}{2}\Lambda_a n(p)=\\\tfrac{1}{2}\bar n(p')[\hat
\Sigma(p')\tfrac{1}{2}\Lambda_a-\tfrac{1}{2}\bar \Lambda_a \hat
\Sigma(p)]n(p)
\end{eqnarray*}
\noindent General strategy of computing the non-Abelian gauge boson
mass matrix is described in detail in \cite{benes2}. It amounts to
computing the matrix effective loop-generated tadpoles between the
'would-be' NG bosons and the flavor gluons. They imply the massless
pole in the effective tree-level longitudinal part of the flavor
gluon polarization tensor. Its residue is the flavor gluon mass
matrix. The apparently non-urgent explicit computation of the flavor
gluon mass matrix along these lines is in progress. For an estimate
of the value of the flavor gluon masses $M_a$ it is quite sufficient
at the moment to neglect the matrix structure, and to use the
original Pagels-Stokar formula \cite{ps}
\begin{equation}
F^2=8N\int\frac{d^4p}{(2\pi)^4}\frac{\Sigma^2(p^2)-\tfrac{1}{4}p^2(\Sigma^2(p^2))^{'}}{(p^2+\Sigma^2)^2}
\label{ps}
\end{equation}
relating the 'would-be' NG boson decay constant $F$ with the
corresponding gauge boson mass $M$ \cite{tc}: $M^2\sim h^2F^2$. Here
$N=3$ is a loop factor. With the explicit form of
$\Sigma(p^2)=(\Lambda^2/p)\gamma\equiv M_{R}^2/p$ at hand the
integral is easily computed, and we have $F^2=\tfrac{15}{16\pi}
M_{R}^2$. Hence,
\begin{equation}
\boxed{M_{aC} \sim M_R}
\end{equation}
Because the sterile neutrino masses are huge there is no problem
with the flavor changing electric charge conserving processes
transmitted by flavor gluons.

Finally, it is easy to show (see (\ref{Pa})) that for $p'\rightarrow
p$ the neutrino-'would-be' NG couplings have the matrix form
\begin{equation}
P_a(p,p)\sim \{\Sigma(p),\tfrac{1}{2}\lambda_a\}\gamma_5,
\phantom{bbbb}a=1,3,4,6,8,\label{Psym}
\end{equation}
and
\begin{equation}
P_a(p,p)\sim [\Sigma(p),\tfrac{1}{2}\lambda_a],\phantom{bbbb}
a=2,5,7\label{Panti}
\end{equation}
We will use these formulas in the following for comparison with the
canonical Higgs mechanism applied to the Lagrangian (\ref{nuR}).

\subsection{3. The Higgs sextet for the sterile neutrino sector}

In the language of the many-body theory we have so far modified
dynamically the dispersion laws of {\it quasi-particles}
corresponding to the Lagrangian (\ref{nuR}). In general, the
quasiparticles are the harmonic oscillator-like excitations created
by the quantum fields (monomials) present in a Lagrangian.
Explicitly, we have generated the masses of neutrinos and of the
flavor gluons by the {\it strong} flavor gluon gauge interaction.

Dynamical generation of flavor gluon masses demands the existence of
specific {\it collective excitations}, the composite 'would-be' NG
bosons. They are guaranteed by the Goldstone theorem and visualized
by the Ward-Takahashi identities. In general, the collective
excitations (bound states) are the excitations created by certain
polynomials of the original quantum fields. Their very formation in
relativistic quantum field theory requires a strong force.

On the other hand, how to generate spontaneously the masses of
sterile neutrino fields and the masses of the $SU(3)_f$ flavor gluon
fields in the Lagrangian (\ref{nuR}) is notoriously known: Simply
one has to add to it an appropriate {\it weakly interacting} Higgs
sector.

In fact, the desired Higgs multiplet is known \cite{bhs}: In the
following we consider the condensing elementary scalar Higgs field
$\Phi_{ij}(x)$ in the complex symmetric sextet representation of
$SU(3)$. Its condensate generates different Majorana masses to all
sterile neutrinos, and at the same time it generates different
masses to all eight flavor gluons. The general Higgs mechanism for
the $U(3)$ symmetry with the Higgs sextet was discussed in entirely
different context of colored superconductors in great detail in
\cite{bhs}. For comparison with the previous Section we briefly
summarize the main steps.

Under the $SU(3)$ the $\Phi$ transforms as a complex symmetric
matrix, $\Phi\rightarrow U \Phi U^T$. The general Higgs Lagrangian
invariant under the $SU(3)\times U(1)$ symmetry has the standard
form
\begin{eqnarray}
{\cal L}_{H}=(D_{\mu}\Phi)^{+}D_{\mu}\Phi-V(\Phi)+{\cal L}_{Y}
\label{LH}
\end{eqnarray}
where
\begin{eqnarray}
D_{\mu}\Phi=\partial_{\mu}\Phi
-ihC^a_{\mu}(\tfrac{1}{2}\lambda_a\Phi+\Phi\tfrac{1}{2}\lambda_a^{T})
\label{D}
\end{eqnarray}
and
\begin{eqnarray}
{\cal L}_{Y}=g_Y \bar \nu_R \Phi (\nu_R)^{{\cal C}} + h.c.
\label{LY}
\end{eqnarray}
The potential $V(\Phi)$ is a function of three independent
invariants $\rm det(\Phi^{+}\Phi)$, $\rm tr(\Phi^{+}\Phi)$, and $\rm
tr(\Phi^{+}\Phi)^2$. It determines the constant (due to the
translational invariance of the vacuum) vacuum expectation value of
the Higgs field $\Phi$:
\begin{eqnarray*}
\phi=\langle \Phi \rangle_0
\end{eqnarray*}
Employing the liberty of writing $\phi$ as $\phi=U \rm v U^{T}$
where $U$ is a convenient unitary matrix, we can cast $\phi$ into a
real, diagonal matrix with non-negative entries, ordered by their
values:
\begin{eqnarray}
\rm v=\rm diag (v_1,v_2,v_3) \label{Delta}
\end{eqnarray}
with $\rm v_1>v_2>v_3>0$. This form of the v.e.v. follows from the
particular form of $V(\Phi)$. Because we argue in terms of an
effective field theory we ignore the fact that this particular form
of $V$ violates the renormalizability (contains the polynomials in
$\Phi$ of the order higher than four).

It is self-evident that the vacuum expectation value (\ref{Delta})
breaks the $SU(3)\times U(1)$ symmetry spontaneously and completely
: The twelve real scalar fields of the complex sextet $\Phi$
decompose as follows: (1) There is one true NG mode $\theta(x)$ of
the spontaneously broken global $U(1)$ symmetry. (2) There are eight
($a=1,...,8$) 'would-be' NG bosons $\theta_a(x)$ which become the
longitudinal components of massive flavor gluons, and disappear from
the physical spectrum. (3) There are three ($i=1,2,3$) massive
radial modes, the Higgs bosons $\chi_i(x)$. Explicitly, and for
small fields we have
\begin{eqnarray}
\Phi=e^{i\tfrac{1}{2}\lambda_a \theta_a}
\tfrac{1}{\surd 2}e^{i\theta}{\rm diag\  (v+\chi)} e^{i\tfrac{1}{2}\lambda_b^{T}\theta_b} \nonumber\\
\approx [{\rm diag\ (v+\chi)} +i\theta \pm i\theta_a \{ {\rm diag\
(v+\chi)},\tfrac{1}{2}\lambda_a\}_{\pm}]\label{sextet}
\end{eqnarray}
where $+$ means the anticommutator and $a=1,3,4,6,8$ while $-$ means
the commutator and $a=2,5,7$.

For most of practical purposes it is convenient to work in the
unitary gauge which eliminates the 'would-be- NG bosons. We set
\begin{eqnarray}
\Phi(x)=\tfrac{1}{\surd 2}e^{i\theta}\rm
diag(v_1+\chi_1,v_2+\chi_2,v_3+\chi_3) \label{PhiU}
\end{eqnarray}
and ignore the small field $\theta(x)$ in considerations which
follow. Due to the axial anomaly its fate is in fact nontrivial, and
will be discussed in Sect.V in some detail.

1. Substitution of (\ref{PhiU}) into the kinetic term in (\ref{LH})
results in two terms: (i) The quadratic polynomial in the fields $C$
defines the mass matrix of flavor gluons,
\begin{widetext}
\begin{multline*}
M^2 = h^2\times \\
\left[
\begin{array}{cccccccc}
(v_1+v_2)^2 & 0 & 0 & 0 & 0 & 0 & 0 & 0\\
0 & (v_1-v_2)^2 & 0 & 0 & 0 & 0 & 0 & 0\\
0 & 0 & 2(v_1^2+v_2^2) & 0 & 0 & 0 & 0 & \frac{2}{\sqrt3}(v_1^2-v_2^2)\\
0 & 0 & 0 & (v_1+v_3)^2 & 0 & 0 & 0 & 0\\
0 & 0 & 0 & 0 & (v_1-v_3)^2 & 0 & 0 & 0\\
0 & 0 & 0 & 0 & 0 & (v_2+v_3)^2 & 0 & 0\\
0 & 0 & 0 & 0 & 0 & 0 & (v_2-v_3)^2 & 0\\
0 & 0 & \frac{2}{\sqrt3}(v_1^2-v_2^2) & 0 & 0 & 0 & 0 &
\frac23(v_1^2+v_2^2+4v_3^2)
\end{array}
\right]
\end{multline*}
\end{widetext}
All masses are nonzero and unequal, with the $(3,8)$ mixing to be
done. For completeness we mention: All Majorana masses come out
huge. The masses of flavor gluons which change flavor must also be
huge. This is guaranteed. Masses of the flavor diagonal flavor
gluons $C_3$ and $C_8$ can be, upon diagonalization, much lighter.
(ii) The higher polynomials define the tree-level interactions of
the Higgs fields $\chi_i$ with flavor gluons $C$.

2. Substitution of (\ref{PhiU}) into the potential $V$ results in
the mass terms of the Higgs fields $\chi_i$ and their
self-interactions.

3. Substitution of (\ref{PhiU}) into the the Yukawa interaction
(\ref{LY}) results in the mass term of the Majorana neutrinos
\begin{eqnarray}
{\cal L}_{M}=\bar n \tfrac{g_Y}{\surd 2}{\rm v}n
\end{eqnarray}
and in the Yukawa interaction of the massive Majorana neutrinos with
the massive Higgs fields $\chi_i$:
\begin{eqnarray}
{\cal L}_{Y}=\tfrac{g_Y}{\surd 2}\bar n {\rm \chi}n
\end{eqnarray}

\subsection{4. A lesson}
The analysis presented above suggests a duality between the {\it
weak coupling canonical Higgs mechanism} of spontaneous generation
of fermion and gauge boson masses and underlying {\it strong
coupling dynamical generation of fermion and gauge boson masses}
without elementary scalar fields. Such a conjecture is supported
also by the paper \cite{kuti}: It argues in favor of an equivalence
of the top quark condensate model (strong coupling) and the SM with
the elementary Higgs field (weak coupling). Without much imagination
but with a lot of reservations our strong-coupling model with
massive flavor gluons can be roughly approximated by the fourfermion
interactions. Similar conjecture is formulated also in a recent
paper \cite{sannino}.

Comparison of the Yukawa couplings (\ref{Psym}, \ref{Panti}) of the
composite 'would-be' NG bosons with the Yukawa couplings of the
elementary 'would-be' NG bosons shown in (\ref{sextet}) reveals that
they are {\it identical}. The former 'would-be' NG bosons are the
bound states not present in the original Lagrangian (\ref{nuR}), but
convincingly predicted on the basis of symmetry considerations by
the WT identity. The latter ones are fully pre-prepared, together
with their massive partners in the Lagrangian (\ref{LH}) in the
elementary Higgs field $\Phi$. The conjecture is that there are
three composite massive $\chi_i$ also in quantum flavor dynamics,
although we are not aware of an existence theorem for them. Their
effective Yukawa interactions with the massive Majorana neutrinos
should be the same as of the elementary ones. Their effective
interactions with the massive flavor gluons should, however, be
derivable by computing the corresponding UV finite neutrino loops.
This apparently non-urgent computation is also in progress.

Can such superheavy scalar particles be useful?  Apparently yes: We
find amusing that one such a superheavy
right-handed-neutrino-composite scalar bound by a NJL four-fermion
interaction was suggested \cite{barenboim} as a candidate for the
inflaton \cite{inflaton}. We point out that there are good reasons
in the literature for considering more inflatons \cite{liddle}. If
the reasons for the existence of namely three of them are
irresistible they could also be good for answering the question why
there are three families \cite{yanagida2}.

We will refer to the duality discussed above  further in the
following: First, to fix the properties of the composite Higgs-like
bosons in Sect.IV. Second, to argue against formation of the
Majorana mass term of the left-handed neutrinos in Sect.III.2.

\section{III. Dynamical electroweak symmetry breaking}

In accord with our strategy the first step is the generation of
fermion masses (chirality-changing fermion self-energies) of the
electroweakly interacting fermions by the strong-coupling quantum
flavor dynamics.

In the second step we demonstrate that the composite electroweak
'would-be' NG bosons resulting from the step one give rise to masses
$m_W, m_Z$ of the $W$ and $Z$ bosons. These masses are expressed in
terms of the fermion masses by sum rules.

\subsection{1. Dynamical generation of charged-lepton and quark masses}

The chirality-changing fermion self-energy is a matrix bridge
between the left-handed and the right-handed fermion field
multiplets with given electric charge. In the Majorana neutrino case
the left- and the right-handed neutrino fields are related by charge
conjugation, and the resulting $\Sigma$ ($3\times 3=\bar 3_a + 6_s$)
is a symmetric matrix by Pauli principle. In the Dirac case the
left- and the right-handed fermion fields are the independent fields
both transforming as triplets
($T_a(L)=T_a(R)=\tfrac{1}{2}\lambda_a$), and the resulting $\Sigma$
is a general complex $\bar 3 \times 3=1+8$ matrix. Its form is found
by solving the SD equation as in the case of the Majorana case.

Clearly, with $SU(2)_L \times U(1)_Y$ electroweak gauge interactions
switched off there is nothing in the model which would distinguish
between $\Sigma$ matrices of different fermion species $f=u, d, e,
\nu$. Consequently the flavor-dependent mass matrices of these
different fermion species must come out equal:
\begin{eqnarray*}
\Sigma_t=\Sigma_b=\Sigma_{\tau}=\Sigma_{\nu_{\tau}} \\
\Sigma_c=\Sigma_s=\Sigma_{\mu}=\Sigma_{\nu_{\mu}}\\
\Sigma_u=\Sigma_d=\Sigma_e=\Sigma_{\nu_e}
\end{eqnarray*}
Hence for $f=u, d, e, \nu$ we get the fermion masses $m_i(f)$
independent of $f$.

In the Appendix the corresponding SD equation is solved (with
fermion mixing neglected), and the resulting mass formula is
\begin{equation}
\boxed{m_i(f)=\Lambda \phantom{b} \rm exp \phantom{b}
(-1/4\alpha_{ii})}\label{bare}
\end{equation}
where
\begin{eqnarray*}
\alpha_{11}&=&\tfrac{3}{64\pi^2}(g_{33}+\tfrac{2}{\sqrt
3}g_{38}+\tfrac{1}{3}g_{88})\\
\alpha_{22}&=&\tfrac{3}{64\pi^2}(g_{33}-\tfrac{2}{\sqrt
3}g_{38}+\tfrac{1}{3}g_{88})\\
\alpha_{33}&=&\tfrac{3}{64\pi^2}\tfrac{4}{3}g_{88}
\end{eqnarray*}

In reality, the Abelian gauge field $B$ of the gauge electroweak
interactions, although interacting identically with different
fermion families, does distinguish between different fermion
species. Different types of chiral fermions differ by having
different weak hypercharges $Y$, given uniquely by different
electric charges. For convenience they are reminded here:
\begin{eqnarray*}
Y(l_L)&=&-1,\phantom{bbb}Y(e_R)=-2,\phantom{bbb}Y(\nu_R)=0\\
Y(q_L)&=&\tfrac{1}{3},\phantom{bbbb}Y(u_R)=\tfrac{4}{3},\phantom{bbbbb}Y(d_R)=-\tfrac{2}{3}
\end{eqnarray*}
Radiative corrections due to the non-Abelian electroweak gauge
fields $A_i$, interacting universally  with all left-handed fermion
fields give rise to a universal contribution and need not be
considered.

Frankly, we do not know at the moment how to implement the $B$
radiative corrections into the separable Ansatz. Clarification of
this important point requires further work. It is nevertheless
justified to consider formally the effective couplings $\alpha_{ii}$
in the fermion mass formula above fermion-type dependent, and write
it in a generic form
\begin{equation}
\boxed{m_i(f)=\Lambda \phantom{b} \rm exp \phantom{b}
(-1/4\alpha_{ii}(f))} \label{mif}
\end{equation}
Here $f=u, d, e, \nu$ and $\alpha_{ii}(f)$ are, ultimately, the
effective couplings $\alpha_{ii}$ with the calculable weak
hypercharge radiative corrections included. A hope is that the
exponential dependence of the fermion masses upon the effective
couplings will describe the observed differences between fermion
masses of fermions with different electric charges.

\subsection{2. Neutrino mass spectrum}

Computation of the neutrino mass spectrum in the present model is
more subtle. It requires the knowledge of the neutrino mass matrix
in its general form of a complex symmetric $6 \times 6$ matrix
\begin{eqnarray}
\Sigma_\nu &=& \left(\begin{array}{cc}
\Sigma_{L}     &  \Sigma_{D} \\
\Sigma_{D}^T  &  \Sigma_{R} \\
\end{array}\right)
\label{Sigmanu}
\end{eqnarray}
Here $\Sigma_{L}$ is the Majorana self-energy of three active
left-handed neutrinos to be discussed, $\Sigma_{D}\equiv \Sigma$ is
the Dirac self-energy of the neutrinos computed in Sect.III.1
(without electroweak corrections), and $\Sigma_{R}$ is the Majorana
self-energy of the triplet of the right-handed neutrinos computed in
Sect.II.1.

What can we say about $\Sigma_{L}$? First of all, three $\nu_L$
belong to three electroweak doublets $l_L^T=(\nu_L, e_L)$. In accord
with our reasoning of Sect.II.4 the spontaneous generation of
$\Sigma_{L}\neq 0$ would imply considering the condensing complex
triplet Higgs field having the quantum numbers of the elementary
scalar Higgs field of the triplet Majoron model \cite{tripletM}. The
$SU(2)_L$ triplet is required by Pauli principle. To postulate the
existence of the field
\begin{equation*}
\phi_a=(\phi^{(0)},\phi^{(+)},\phi^{(++)})
\end{equation*}
does not cost anything provided it is elementary. Whether such a
field has the right to exist as a bound state of two $l_L$s is a
complicated issue of the underlying strong-coupling quantum flavor
dynamics, definitely outside the scope of the present paper. Here we
merely wishfully assume that the formation of a doubly charged
composite Higgs-like field would be energetically very costly
because of the Coulomb repulsion, and the complex composite triplet
will not be formed. Consequently, there would be no condensate
$\Sigma_{L}$ and we conclude that the neutrino mass eigenstates are
determined by the neutrino mass matrix of the famous seesaw
\cite{seesaw} form
\begin{eqnarray}
\Sigma_\nu &=& \left(\begin{array}{cc}
0    &  \Sigma_{D} \\
\Sigma_{D}^T  &  \Sigma_{R} \\
\end{array}\right)
\label{Sigmanua}
\end{eqnarray}
This implies that, upon diagonalization, there are three Majorana
neutrinos with huge masses $\sim M_R$, and three active light
Majorana neutrinos with masses
\begin{equation}
\boxed{m_{\nu} \sim m^2_D/M_R}
\end{equation}\label{mnu}
Clearly, for any dynamics pretending to compute the fermion masses
the prediction of the neutrino mass spectrum is a crystalline
challenge. Today, masses of the electrically charged leptons and
quarks can 'only' be post-dicted. Referring to the simple analysis
presented in the Appendix it seems that the masses $M_R$ cannot be
made arbitrarily large: If we want to keep all $g_{ab}$ of the
similar order of magnitude, we are restricted by the fact that those
$g$ giving rise to the Dirac fermion masses are essentially fixed.
Consequently, the masses of three active Majorana neutrinos cannot
be apparently arbitrarily small. Other possibility, also mentioned
in the Appendix, is that the Majorana and Dirac fermion masses are
generated at different scales.

\subsection{3. Dynamical generation of intermediate gauge boson masses}

Fermion masses of the electroweakly interacting fermions or, more
generally, their chirality-changing fermion proper self-energies
$\Sigma(p^2)$, generated nonperturbatively by the strong gauge
flavor dynamics, break spontaneously the electroweak $SU(2)_L \times
U(1)_Y$ gauge symmetry down to unbroken electromagnetic $U(1)_{em}$.
Consequently, the $W,Z$ gauge bosons must acquire masses
proportional to $\Sigma$.

It is very important that the gauge electroweak $SU(2)_L \times
U(1)_Y$ tie together in a unique way many otherwise independent
global $SU(2)_L$ and $U(1)_{L,R}$ symmetries created by the chiral
currents of different fermion species. With electroweak gauge
interactions switched off there would be plenty of
phenomenologically unacceptable real NG bosons. With electroweak
interactions switched on there are just three composite
multi-component 'would-be' NG bosons which give rise to the masses
of $W$ and $Z$ bosons.

We proceed as in the case of the dynamical generation of flavor
gluon masses described in Sect.II.2. There we acted as if the
fermion-gauge boson coupling $h$ were small. Here the couplings $g,
g'$ are indeed small.

Consider the vertex parts $\Gamma^{\alpha}_{fW}(p+q,p)$ and
$\Gamma^{\alpha}_{fZ}(p+q,p)$ following from the electroweak WT
identities \cite{huang-mendel} for one, heaviest, quark doublet
$f=(t,b)$. In our simplified world without mixing and without
perturbative contributions from the electroweak interactions the
self-energies are equal,
$\Sigma_t(p^2)=\Sigma_b(p^2)=\Sigma_{\tau}(p^2)=\Sigma_{\nu_{\tau}}(p^2)=\Sigma_f(p^2)$.

\begin{multline*}
\Gamma^{\alpha}_{W}(p+q,p)=\frac{g}{2\sqrt2}\{\gamma^{\alpha}(1-\gamma_5)-\\
-\frac{q^{\alpha}}{q^2}[(1-\gamma_5)\Sigma_f(p+q)-(1+\gamma_5)\Sigma_f(p)]\},
\end{multline*}
\begin{multline*}
\Gamma^{\alpha}_{Z}(p+q,p)=\frac{g}{2\cos\theta_W}\{t_3\gamma^{\alpha}(1-\gamma_5)-\\
-2Q\gamma^{\alpha}\sin^2\theta_W-\frac{q^{\alpha}}{q^2}t_3
[\Sigma_f(p+q)+\Sigma_f(p)]\gamma_5\}.
\end{multline*}

Further steps are standard (see, e.g.,\cite{benes3}). We extract
from the pole terms of the WT identities the effective
fermion-'would-be' NG boson vertices $P$
\begin{equation*}
P_{\pm}^f=\frac{1}{4F}[(1\mp\gamma_5)\Sigma_f(p+q)-(1\pm\gamma_5)\Sigma_f(p)]
\end{equation*}
\begin{equation*}
P_0^f=\frac{1}{4F}\gamma_5t_3(\Sigma_f(p+q)+\Sigma_f(p)),
\end{equation*}
and the effective f-component of the 'would-be' NG bilinear
couplings with the gauge fields $J^{\mu}$:
\begin{multline*}
J_{fW}^{\mu}(q)=\Tr\int\frac{d^4k}{(2\pi)^4}P_{-}S_f(k+q)
\frac{g}{2\sqrt2}\gamma^{\mu}(1-\gamma_5)S_f(k)
\end{multline*}
\begin{multline*}
J_{fZ}^{\mu}(q)=\Tr\int\frac{d^4k}{(2\pi)^4}P_0S_f(k+q)\frac{g}{2cos\theta_W}\\
[t_3\gamma^{\mu}(1-\gamma_5)-2Q\gamma^{\mu}sin^2\theta_W]S_f(k)
\end{multline*}
Here $F$ is the normalization factor to be computed explicitly in
terms of all $\Sigma_f$ in the model.

Without fermion mixing the Yukawa $u,d$ quark-'would-be NG boson
vertices have for $q^{\mu}=0$ and $\Sigma(p^2=m^2)=m$ the form
\begin{equation}
P_{+}^f=\frac{m_f}{2F}\gamma_5\label{pplus}
\end{equation}
\begin{equation}
P_0^f=\frac{m_f}{2F}t_3\gamma_5\label{pzero}
\end{equation}

The effective vertices $P$ will be employed in the next Section for
the identification of the form of the operator of the composite
Higgs field. The quantities $J^{\mu}$ guarantee the massless pole in
the longitudinal part of the gauge boson polarization tensors i.e.,
their masses. Here we compute them again using the Pagels-Stokar
formula as it was done in technicolor \cite{tc}:
\begin{equation}
F_f^2=8N\int\frac{d^4p}{(2\pi)^4}\frac{\Sigma_f^2(p^2)-\tfrac{1}{4}p^2(\Sigma_f^2(p^2))^{'}}{(p^2+\Sigma_f^2)^2}
\label{ps}
\end{equation}
As before, $F_f^2=\tfrac{5}{16\pi}N m_f^2$. Here $N=3+1=4$ where $3$
stands for three colors of the $t,b$ colored quarks, and $1$ stands
for $\tau,\nu_{\tau}$ colorless leptons. Because the composite
'would-be' NG bosons made of all electroweak doublets and their
corresponding singlets incoherently contribute, we have
\begin{equation*}
F^2=\sum_f F_f^2
\end{equation*}
Unlike the flavor gluon masses the numerical values of the
electroweak gauge bosons $W, Z$ are the very important parameters of
the world at present energies. Considering only the heaviest
fermions with the common mass $m$, we have
\begin{equation*}
F^2=\tfrac{5}{16\pi}N m^2=\tfrac{5}{4\pi}m^2
\end{equation*}
In general,
\begin{equation}
\boxed{m_W^2=\tfrac{1}{4}g^2\tfrac{5}{4\pi} \sum_f m_f^2}\label{mw}
\end{equation}
\begin{equation}
\boxed{m_Z^2=\tfrac{1}{4}(g^2 + g'^2)\tfrac{5}{4\pi} \sum_f
m_f^2}\label{mz}
\end{equation}
At the present exploratory stage not taking into account either the
electroweak corrections or the fermion mixing the mass $m$ is not
known. Expecting its value of order of the electroweak scale $v$ we
conclude that the result is satisfactory. It also should be
remembered that the relation between intermediate vector boson
masses and the masses of electroweakly interacting fermions is
sensitive to the explicit form of $\Sigma(p^2)$.

\section{IV. The CERN Higgs and its two relatives}

So far we have shown how the strong gauge flavor dynamics generates
the calculable masses of leptons and quarks and demonstrated how, as
a consequence of the existence theorem, it generates the composite
'would-be' NG bosons giving masses to the intermediate electroweak
bosons $W,Z$.

Does the strong quantum flavor dynamics produce also composite Higgs
particles ? As far as we can see, there is no existence theorem for
such states. We will follow the logic of Sect.II and search for the
fermion-antifermion composite operator which transforms according to
a representation of the gauge group, and as the necessary
consequence it contains the composite 'would-be' NG bosons found
previously by the analysis of the WT identities. The remaining
partner(s) will be identified with the composite Higgs(es).

We start with the electroweak symmetry. As a necessary condition the
composite scalar operator constructed from the covariantly
tranforming fermion fields should contain three components of the
'would-be' NG bosons $\pi$, which in the elementary scalar Higgs
field are pre-prepared in the decomposition
\[
\Phi(x)= \rm exp(\frac{i}{v}\pi^a(x)  \tau^a) \left(
\begin{array}{cc}
0\\
\tfrac{1}{\sqrt 2}(v+h(x))\end{array} \right)\equiv \left(
\begin{array}{cc}
\phi^{+}\\
\phi^{0} \end{array} \right)
\]
For small fields we have
\begin{eqnarray*}
\phi^{+}&=&\tfrac{1}{\sqrt 2}(\pi_2+i\pi_1)\\
\phi^{0}&=&\tfrac{1}{\sqrt 2}(v+h-i\pi_3)
\end{eqnarray*}
Here $h$ is the physical Higgs field and $v=246\phantom{b} \rm GeV$
is the electroweak condensate.

\subsection{1. The Yukawa couplings of the Higgs boson}

In the Standard model the quark masses $m_u, m_d$ are generated from
the $SU(2)_L \times U(1)_Y$ invariant Yukawa interaction
\begin{equation}
{\cal L}_Y=y_u\bar q_L u_R \tilde \Phi + y_d \bar q_L d_R \Phi +
h.c.\label{lY1}
\end{equation}
where the charge conjugate Higgs field is $\tilde \Phi\equiv i\tau^2
\Phi^{*}$. In our approximation $m_u=m_d=y_u v/\sqrt 2 = y_u v/\sqrt
2=m$ we get
\begin{eqnarray*}
{\cal L}_Y=m(\bar u u + \bar d d) + \frac{m}{v}(\bar u u + \bar d
d)h \phantom{bbbbbbb}\\ + \{\frac{m}{2v}\bar u \gamma_5 d (\pi_2 +
i\pi_1) + h.c.\}+\frac{m}{2v}i[\bar u \gamma_5 u - \bar d \gamma_5
d]\pi_3
\end{eqnarray*}
Standard interpretation of ${\cal L}_Y$ is the following: In the
unitary gauge, in which all $\pi_i$ vanish, ${\cal L}_Y$ turns into
the ordinary fermion mass term plus parity-conserving Yukawa
interaction of the real $0^{+}$ field $h$ with couplings
proportional to the fermion masses.

Comparison of ${\cal L}_Y$ with our results obtained so far implies
the following: First, we also generate spontaneously the fermion
masses. While the Higgs mechanism is a tree-level effect, ours is
genuinely quantal. Second, the terms in ${\cal L}_Y$ containing the
elementary 'would-be' NG bosons should be compared with the
composite ones, signalled in the formulas (\ref{pplus}) and
(\ref{pzero}) by their massless poles. Clearly, the correspondence
is complete, provided
\begin{equation}
F=v
\end{equation}
Third, we believe that this coincidence justifies a conjecture that
also in our model there is a composite {\it multicomponent} $h$ as a
massive partner of the NG bosons. It is seen as the real neutral
component of the composite operator
\begin{equation*}
\Phi=\sum(\Phi(\nu)+\tilde \Phi(e) + \Phi(d)+\tilde \Phi(u))
\end{equation*}
where in the quark sector
\[
\Phi(d)=\frac{1}{F^2} \bar d_R q_L = \frac{1}{F^2} \left(
\begin{array}{cc}
\bar d_R u_L \\ \bar d_R d_L
\end{array}
\right)
\]
\[
\Phi(u)=\frac{1}{F^2} \bar u_R q_L = \frac{1}{F^2} \left(
\begin{array}{cc}
\bar u_R u_L \\ \bar u_R d_L
\end{array}
\right)
\]
The leptonic composite doublets are constructed analogously.

Consequently, the Yukawa vertices of the composite $h$, which
eventually enter the fermion loops, have the form
%\begin{widetext}
\begin{equation*}
{\cal L}_Y=\frac{h}{F}\sum[(\bar u_L \Sigma u_R + \bar u_R \Sigma
u_L) + (\bar d_L \Sigma d_R + \bar d_R \Sigma d_L)]
\end{equation*}
%\end{widetext}
Presence of fermion proper self-energies $\Sigma(p^2)$  provides
necessary softening of the corresponding integrals at high momenta.
The sum is over all upper (u) and lower (d) fermions (both quarks
and leptons) in the electroweak doublets.

\subsection{2. Gauge couplings of the Higgs boson}

Phenomenologically the most important question to be answered is how
the composite Higgs $h$ interacts with the electroweak gauge bosons
$W$, $Z$ and $A$. Without any computations it is obvious that the
resulting interactions must generically differ from the Standard
model ones: In the present model all electroweak gauge fields
interact directly only with the chiral lepton and quark fields.
Consequently, {\it all electroweak gauge fields interact with the
composite $h$ merely via the UV finite fermion loops}.

In contrast, the SM gauge interactions of the $W,Z$ fields with $h$
are the tree-level ones coming from the covariant derivative of the
complex doublet Higgs field $\Phi$:
\begin{eqnarray*}
{\cal
L}_{h,W,Z}=\frac{1}{8}(2vh+h^2)[2g^2W^{-}_{\mu}W^{+\mu}+(g^2+g'^2)Z_{\mu}Z^{\mu}]\\
=+gm_W W^{-}_{\mu}W^{+\mu}h + \frac{1}{2 \rm cos
\theta_W}gm_ZZ_{\mu}Z^{\mu}h\\
+\frac{1}{4}g^2W^{-}_{\mu}W^{+\mu}h^2+\frac{g^2}{8 \rm cos^2
\theta_W}Z_{\mu}Z^{\mu}h^2 \label{lhw}
\end{eqnarray*}
The photon interacts in SM with the Higgs field via the fermion and
the $W$ loops. A nasty remark might be that in this respect the
unification of weak and electromagnetic interactions in the Standard
model is rather strange. The detailed derivation of the effective
interactions deserves separate work now in progress.

\subsection{3. The Higgs boson mass}

Finally, there is a question of the Higgs boson mass. In the
Standard model the tree-level answer is exceedingly simple:
\begin{equation}
m_h^2=2\lambda v^2 \label{mh}
\end{equation}
where $\lambda$ is the perturbative quartic Higgs field
self-coupling.

We do not know how to compute reliably the mass of the non-NG-type
collective excitation $h$ in our strong-coupling model. Referring to
the similarity of our approach with the top quark condensate model
(BHL)\cite{bhl} we weaken their strong coupling result for the Higgs
boson mass $m_h=2m_{\rm top}$ into an estimate $m_h\sim O(F)$.

\subsection{4. Two Higgs-boson relatives}

Clearly, the dynamically generated masses of the electroweakly
interacting leptons and quarks break spontaneously not only the
electroweak symmetry but also the flavor symmetry $SU(3)_f$. More
precisely, if the fermion self-energy $\Sigma$ (fermion mass) is
written as
\begin{eqnarray*}
\Sigma&=&\Sigma_0 \lambda_0+\Sigma_3 \tfrac{1}{2}\lambda_3+\Sigma_8 \tfrac{1}{2}\lambda_8\\
&\equiv&\Sigma_0 \lambda_0+\Sigma'
\end{eqnarray*}
($\lambda_0=\sqrt{\tfrac{2}{3}} \textbf{1}$)) then only
$\Sigma'=\Sigma_3 \tfrac{1}{2}\lambda_3+\Sigma_8
\tfrac{1}{2}\lambda_8$ is $SU(3)_f$ symmetry-breaking, as mentioned
in the Introduction. The corresponding composite 'would-be' NG
bosons are visualized as massless poles by virtue of the $SU(3)_f$
WT identity for the vertex valid for arbitrary flavor triplet
fermion $f=u,d,e,\nu$
\begin{eqnarray*}
\Gamma^{\alpha}_{aC}(p+q,p)=h\{\tfrac{1}{2}\lambda_a-\\
-\frac{q^{\alpha}}{q^2}[\Sigma(p+q)\tfrac{1}{2}\lambda_a-\tfrac{1}{2}\lambda_a\Sigma(p)]\}
\end{eqnarray*}
As before we can extract from the pole part of $\Gamma$ two
quantities \cite{benes2}:

First, the effective bilinear flavor gluon-'would-be' NG couplings
(vectorial tadpoles) giving rise to small contributions of order $m$
to the huge masses of flavor gluons. Detailed analysis of mixing of
these composite 'would-be' NG bosons with the most important sterile
neutrino-antineutrino composite ones requires extra work.

Second, the Yukawa couplings of fermions with the composite
'would-be' NG bosons $\theta_a$ are proportional to
\begin{equation}
P_a \sim -[m',\tfrac{1}{2}\lambda_a]\label{P}
\end{equation}
(we set $\Sigma(p^2=m^2)=m)$).

Notice that in the commutator the term $m_0$ proportional to the
unit matrix, commuting with all eight $\lambda$ matrices, is absent.
Consequently, there are six composite 'would-be' NG bosons
corresponding to the generators (1,2,4,5,6,7)(coupled with fermions
$f$ in a uniquely prescribed way ($m'=m_{3}
\tfrac{1}{2}\lambda_3+m_{8} \tfrac{1}{2}\lambda_8$).

Our task is now to find such an ordinary Higgs $SU(3)_f$ multiplet
of the elementary spinless fields $\phi$ the condensate of which
spontaneously generates $m'$ and at the same time contains six
'pre-prepared' elementary 'would-be' NG bosons coupled to fermions
as in (\ref{P}). Comparison with the dynamical picture described
above then should yield the prediction of the composite Higgs-like
particles and the form of their Yukawa couplings.

It is known \cite{kibble},\cite{kim} that the octet $\phi_a$,
$a=1,...,8$ breaks spontaneously the gauge $SU(3)$ symmetry by its
vacuum condensates $\langle \phi_3\rangle$, $\langle \phi_8\rangle$
down to the unbroken $U(1) \times U(1)$ subgroup. To establish the
correspondence with (\ref{P}) we proceed heuristically as follows
\cite{kim}:

Consider eight small fields in the polar decomposition
\begin{eqnarray}
\Phi(x)=exp [i\theta_a \tfrac{1}{2}\lambda_a](m'+s'(x)) exp
[-i\theta_b \tfrac{1}{2}\lambda_b]\nonumber
\\
\sim m'+s'(x)-i[m',\theta_a \tfrac{1}{2}\lambda_a]\label{phi}
\end{eqnarray}
Simple inspection of the commutator reveals that the six fields
\begin{eqnarray*}
-m_3\theta_2,+m_3\theta_1,-\tfrac{1}{2}(m_3+\sqrt{3}m_8)\theta_5,
\tfrac{1}{2}(m_3+\sqrt{3}m_8)\theta_4,\\
\tfrac{1}{2}(m_3-\sqrt{3}m_8)\theta_7,-\tfrac{1}{2}(m_3-\sqrt{3}m_8)\theta_6
\end{eqnarray*}
are the 'would-be' NG bosons which disappear from the physical
spectrum and contribute to the masses of six flavor-changing flavor
gluons.

From the correspondence between the formulas (\ref{P}) and
(\ref{phi}) we conclude: First, both the strong flavor dynamics and
the elementary scalar octet generate the fermion mass term $m'$.
While the dynamical fermion mass generation is a nonperturbative
quantum loop effect, the use of the elementary scalar octet amounts
merely to a tree-level vacuum condensation. Second, both approaches
reveal in the intermediate state the existence of identically
coupled 'would-be' NG bosons. Third, and most important, we predict
from this correspondence  the existence of two composite scalars
$h_3(x)$ and $h_8(x)$ with peculiar Yukawa interactions with the
electroweakly interacting fermions $f$ of the form
\begin{eqnarray*}
{\cal L'}_Y=\tfrac{m_3}{N}\bar f(x) \tfrac{1}{2}\lambda_3 f(x)
h_3(x)+\tfrac{m_8}{N}\bar f(x) \tfrac{1}{2}\lambda_8 f(x) h_8(x)
\end{eqnarray*}
In loops the hard fermion masses should be replaced by the
corresponding momentum dependent self-energies $\Sigma(p)$. Here $N$
is a normalization factor analogous to the scale $F$ of the previous
section i.e., of the order of the electroweak scale.

Detailed elaboration of the form of the effective interactions of
these scalars with the electroweak gauge particles needs further
work now in progress. The presence of the flavor matrices
$\lambda_{3,8}$ in the Yukawa couplings inevitably implies that the
fermion loop can only be nonzero (and finite) with the fermion mass
insertion proportional to $\Sigma_{3,8}(p)$.

\section{V. Fate of global Abelian chiral symmetries}

There are six Abelian symmetries generated by the charges of six
chiral fermion currents $j^{\mu}_i, i=q_L, u_R, d_R, l_L, e_R,
\nu_R$. Taking into account {\it the quantum effects of axial
anomalies} with four gauge forces in the game we have \cite{thooft}
\begin{eqnarray*}
\partial_{\mu}j_{q_L}^{\mu} & = & \partial_{\mu}(\bar q_L \gamma^{\mu} q_L)  =  -A_Y-9A_W-6A_G-6A_F\\
\partial_{\mu}j_{u_R}^{\mu} & = & \partial_{\mu}(\bar u_R \gamma^{\mu} u_R)  =  8A_Y+3A_G+3A_F \\
\partial_{\mu}j_{d_R}^{\mu} & = & \partial_{\mu}(\bar d_R \gamma^{\mu} d_R)  =  2A_Y+3A_G+3A_F \\
\partial_{\mu}j_{l_L}^{\mu} & = & \partial_{\mu}(\bar l_L \gamma^{\mu} l_L)  =  -3A_Y-3A_W-2A_F \\
\partial_{\mu}j_{e_R}^{\mu} & = & \partial_{\mu}(\bar e_R \gamma^{\mu} e_R)  =  6A_Y+A_F \\
\partial_{\mu}j_{\nu_R}^{\mu} & = & \partial_{\mu}(\bar\nu_R\gamma^{\mu}\nu_R)  =  3A_F
\end{eqnarray*}
Here $A_X=\tfrac{g_X{2}}{32\pi^2}F_X\tilde F_X$, and $X$ abbreviates
the gauge forces $U(1)_Y$, $SU(2)_L$, $SU(3)_c$ and $SU(3)_f$,
respectively.

\subsection{1. Anomaly-free currents}
There are two linear combinations of the currents $j^{\mu}_i$ which
are anomaly-free. They can be parameterized by two real parameters
$e,f$: $j^{\mu}_{e,f} =
-\tfrac{1}{6}(e+3f)j_{q_L}^{\mu}+\tfrac{1}{3}(-2e+3f)j_{u_R}^{\mu}+\tfrac{1}{3}(e-6f)j_{d_R}^{\mu}+
\tfrac{1}{2}(e+3f)j_{l_L}^{\mu}+ej_{e_R}^{\mu}+fj_{\nu_R}^{\mu}$.

For $f=0$, $e=-2$ we get the anomaly free current of the {\it
gauged} weak hypercharge $Y$, $j^{\mu}_Y$. Because the electric
charge $Q$ is $Q=I_3 + \tfrac{1}{2}Y$, the hypercharge current is
hidden in the vectorial electromagnetic current coupled to the
massless photon field $A$, and does not create any 'would-be' NG
boson.

For definiteness we fix quite arbitrarily the other anomaly free
current $j^{\mu}_{Y'}$ by $f\neq0, e=0$:
\begin{eqnarray*}
\tfrac{1}{f}j^{\mu}_{Y'} =
-\tfrac{1}{2}j_{q_L}^{\mu}+j_{u_R}^{\mu}-2j_{d_R}^{\mu}+
\tfrac{3}{2}j_{l_L}^{\mu}+j_{\nu_R}^{\mu}
\end{eqnarray*}
It creates the true fermion-antifermion massless composite NG boson.
Because of its $\nu_R$ component its couplings with fermions are
tiny. To prevent any conflict with data we better gauge the
corresponding Abelian symmetry. We can, because it is anomaly-free.
The NG boson becomes 'would-be' and the new $Z'$ very heavy, with
mass mass $m_{Z'} \sim g'' M_R$ where $g''$ is a new gauge coupling
constant.

\subsection{2. The axions}

One of the four anomalous linear combinations can be chosen as the
{\it vectorial} baryon current
$j^{\mu}_B=\tfrac{1}{3}(j^{\mu}_{q_L}+j^{\mu}_{u_R}+j^{\mu}_{d_R})$
which does not create any pseudo NG boson.

We are left with three anomalous Abelian chiral currents which
create three pseudo NG bosons. Their masses are assumed to be due to
the nonperturbative effects of three non-Abelian forces present in
the game \cite{inst}. It is natural to identify two of them with
those already known in the literature: First is the famous
Weinberg-Wilczek axion $a$ \cite{ww}, massive due to the instanton
of the confining QCD. Second is the Anselm-Uraltsev 'arion' $b$
\cite{au} with the mass expectedly associated with nonperturbative
effects of the electroweak $SU(2)_L$ gauge fields \cite{inst}. The
third one is the new axion $c$ with mass expectedly associated with
nonperturbative effects of the q.f.d. $SU(3)_f$ gauge fields
\cite{inst}. We are not aware of any generally accepted formula for
$m_b$ and $m_c$. Referring to \cite {inst} we merely expect that
they contain the suppression factor due to the screened instantons.

Because the lack of data there is much freedom in fixing the
coefficients in the linear combinations $j_a^{\mu}=\sum_i a_i
j^{\mu}_i$, $j_b^{\mu}=\sum_i b_i j^{\mu}_i$, and $j_c^{\mu}=\sum_i
c_i j^{\mu}_i$, $i=q_L, u_R, d_R, l_L, e_R, \nu_R$. For definiteness
we can simulate the data by demanding: 1. The axion $a$ is
invisible, hence $a_{\nu_R}\neq 0$. 2. The strongest gauge
interaction that gives rise to its mass is QCD. Hence,
$\partial_{\mu} j_a^{\mu}$ must not contain $A_F$. 3. The axion $b$
becomes massive due to the electroweak interactions. Hence,
$\partial_{\mu} j_b^{\mu}$ must not contain $A_F, A_G$. 4. The axion
$c$ does not interact with gluons. Hence, $\partial_{\mu} j_c^{\mu}$
must not contain $A_G$. 5. The requirement $h_{\nu_R}=0$ implies
that the interactions of $c$ with fermions are not suppressed by
$M_R$.

Explicit evaluation of the axion properties requires separate work.
For example, the axions as described above are not the mass
eigenstates, and in the present scheme there should be the axion
mixing and, possibly the axion oscillations. Because the enormous
hierarchy of scales the tiny mixing should be negligible.

{\it The form of the pseudo NG currents fixes the effective
interactions of the pseudo NG bosons.}  1. The WT identities
associated with the pseudo NG currents determine the effective
Yukawa couplings of the pseudo NG bosons with fermions. In separable
approximation they are pseudoscalar. 2. The divergences of the
pseudo NG currents fix the effective interactions of the pseudo NG
bosons $a, b, c$ with the respective gauge fields. There is no way
how they could interact with the gauge particles by the
renormalizable interactions.

{\it The effective interactions of the pseudo NG bosons with
fermions and non-Abelian gauge bosons give rise to their masses}
\cite{ww}. For the axion $a$ massive due to the instanton of the
confining QCD we use the generally accepted estimate \cite{ww} $m_a
\sim \Lambda_{QCD}^2/\Lambda \sim 10^{-2} \rm eV$ from which we fix
the scale $\Lambda \sim 10^{10} \rm GeV$. (ii) For an estimate of
masses of the axions $b$ and $c$ associated with the dynamically
massive non-Abelian gauge sectors $SU(2)_L$ and $SU(3)_f$,
respectively, we are not aware of any generally accepted formula
\cite{ringwald1}. Here we simply wishfully assume that the masses
$m_b$ and $m_c$ of the axions $b$ and $c$ are such that they explain
some astrophysical puzzles currently discussed \cite {ringwald}.

\section{VI. Specific consequences of the model}

We have argued that the $SU(3)_f \times SU(2)_L \times U(1)_Y$
dynamics basically reproduces all observed properties of the
Standard model: First, it yields, at least in our approximation, the
lepton and quark masses. Second, as a {\it necessary} consequence of
spontaneous fermion mass generation the composite 'would-be' NG
bosons inevitably give rise to masses of the electroweak gauge
bosons $W$ and $Z$ proportional to the fermion masses. Third, a {\it
natural} consequence of the existence of three electroweak composite
'would-be' NG bosons is that they have their genuine massive
partner, the composite Higgs-like boson $h$. What are the main
specific consequences of this very rigid model ?

1. The Higgs-like $0^{+}$ boson $h$ is a fermion-antifermion
composite. Although the Yukawa interactions of $h$ with leptons and
quarks are the same as in the Standard model, its interactions with
the electroweak gauge bosons $W, Z, A$, being all loop-generated,
are {\it different}. Experimental data on these interactions provide
the crucial test of the present model.

2. The model predicts {\it two} massive flavored Higgs-like
fermion-antifermion $0^{+}$ scalars $h_{3}$ and $h_{8}$. These are
the massive partners of the composite 'would-be' NG bosons
inevitably following spontaneous breakdown of flavor $SU(3)_f$ by
dynamically generated masses of the electroweakly interacting
leptons and quarks. Both the tree-level Yukawa couplings of $h_{3}$
and $h_{8}$ with fermions and their loop-generated effective
interactions with the electroweak gauge bosons are uniquely fixed.

3. For purely theoretical reason of anomaly freedom we were enforced
to add to the list of the observed SM chiral lepton and quark fields
one triplet of sterile right-handed neutrino fields. This by itself
is most welcome: First, the interactions of sterile neutrinos with
flavor gluons cause the complete dynamical self-breaking of
$SU(3)_f$. Second, we believe that the $SU(3)_f$ with sterile
right-handed neutrinos provides the origin of the seesaw mechanism:
There is a hopefully good reason why the left-handed neutrino
Majorana masses are not dynamically generated. As a consequence the
huge Majorana masses of the right-handed neutrinos are responsible
for the observed lightness of active neutrinos. The observed three
active neutrinos are the extremely light Majorana particles.

4. Mixing of  superheavy sterile Majorana neutrinos implies new
complex phases and therefore a {\it new source of CP violation} in
the model. An extra source of CP violation seems indispensable for
understanding the baryon-antibaryon asymmetry of the Universe
\cite{leptogenesis}.

5. Global anomalous Abelian chiral symmetries of the microscopic
Lagrangian, spontaneously broken by the dynamically generated
fermion masses result in three pseudo-NG bosons. (i) The axions are
the well motivated candidates for dark matter \cite {sikivie}. (ii)
The axions naturally solve several astroparticle puzzles
\cite{ringwald}. (iii) The Weinberg-Wilczek axion naturally solves
the problem of strong CP violation \cite{ww}.

6. The $SU(3)_f$ deals ultimately with one parameter, the the scale
$\Lambda$. The electroweak sector adds to the list of parameters the
unquantized weak hypercharges. These are, however, completely fixed
in the SM by the lepton and quark electric charges, and the number
of free parameters therefore does not increase. There should be
plenty of relations between masses, mixing matrices and gauge
couplings which can be tested. At the present fragmentary
understanding of the strong-coupling infrared flavor dynamics the
mass relations belong more to the realm of dreams rather than to
reality. But sound dreaming \cite{weinberg} is healthy
\cite{richter2}, and we mention merely the sum rules for the gauge
boson masses $m_W, m_Z$ in terms of the masses of electroweakly
interacting fermions.

7. The model contains {\it three} superheavy spin-zero scalars
$\chi_i$ composed of sterile neutrinos with fixed interactions. It
is amazing that such particles can be phenomenologically useful in
cosmology \cite{barenboim}.

8. There should be new bound states with masses of order $\Lambda$.
Although the scale $\Lambda$ of $SU(3)_f$ is very high, there is a
possibility to look for traces of the new gauge flavor dynamics
directly at the extremely high energies \cite{dietrich} in the air
showers.

\section{VII. Conclusions and outlook}

Spontaneous generation of lepton and quark masses in the Standard
model does not provide any understanding of their values: Fermion
masses come out as the Higgs field condensate $v = 246\phantom{b}
\rm GeV$ multiplied by independently renormalized i.e., {\it
theoretically arbitrary, vastly different}, Yukawa couplings. This
is the phenomenological description of fermion masses by
construction. If the recently discovered spinless $125\phantom{b}
\rm GeV$ boson were indeed the Higgs boson of the Standard model
such a sad state of affairs would stay for ever.

We have suggested in this paper to replace the essentially classical
Higgs sector of the SM by a new non-Abelian genuinely quantum
dynamics defined by properly gauging the flavor (family, generation,
horizontal) $SU(3)_f$ index. We have argued that the $SU(3)_f$ gauge
quantum flavor dynamics in its strong coupling regime, due to its
sterile neutrino sector, is not confining but it self-consistently
completely self-breaks. If true this implies that both its
quasi-particle oscillator-type excitations as well as its
bound-state collective excitations have masses which are the
calculable multiples of $\Lambda$. This is the ultimate reason for
our suggestion. The computations of particle masses presented here
are, however, still rather illustrative. Although the Ansatz for the
kernel of the SD equation is very crude, it nicely illustrates the
important point: There are no large and small numbers in the
microscopic Lagrangian. They come out only {\it in solutions} of the
field equations. This happens as a robust, {\it natural}
non-perturbative phenomenon, and should not be called fine tuning.
Derivation of the effective couplings $g_{ab}$ is a dream.

In old days there was nothing wrong with the Higgs sector of the
Standard model from the theory point of view: What could be better
than a renormalizable weak coupling theory ! Latter objection of
'unnaturalness' was always considered by some \cite{richter} as
unwarranted. With the triumphant discovery of the Higgs boson this
theory is now very successful also phenomenologically, and has to be
taken truly seriously, even though as an {\it incomplete effective
field theory}.

Such a view does not preclude attempts at revealing an underlying
microscopic dynamics. The experimentally confirmed properties of the
canonical Higgs mechanism must be, first of all, reproduced by it.
Its phenomenological parameters, e.g. its Yukawa couplings should,
however, be the calculable numbers.

For some (including the author) the guiding idea in attempts
\cite{ho} to find the microscopic dynamics underlying the Higgs one
always was the microscopic theory of superconductivity of Bardeen,
Cooper and Schrieffer (BCS), known to underlie the phenomenological
theory of Ginzburg and Landau (GL): First of all, there is nothing
wrong with the GL theory. It is so beautiful and so deep that it
lead to the prediction of two phenomena awarded by Nobel prizes: The
Josephson effect \cite{joseph}, and the type-II superconductivity
\cite{abri}. We cannot be sure that the potential of the Standard
model with its general Higgs sector, now not restricted by
renormalizability, has already been fully explored. We are inclined
to argue that the Weinberg's famous dimension-five coupling
\cite{wein} should be an integral part of the SM effective
Lagrangian today. It implies the firm {\it prediction}: {\it The
three active neutrinos are the massive Majorana particles.}

The microscopic BCS superconductivity of course reproduces all good
features of the phenomenological GL theory which, under certain
assumptions, can be derived from it \cite{gorkov} . This is the
necessary condition. The validity of BCS is, however, truly tested
where GL has nothing to say: By measuring the dependence of the {\it
electronic} specific heat on temperature below the superconducting
critical temperature $T_c$. In accordance with data this dependence
is exponential due to the gap in the quasi-electron dispersion law.
What is the analogous {\it decisive} test of the dynamics of the
electroweak symmetry breaking ? In the canonical Higgs model the
Higgs condensate exists for ever. In quantum flavor dynamics the
fermion masses are generated only below $\Lambda$. Safe but
impractical way to test the origin of the massiveness of fermions
and of intermediate bosons is to go to very high energies.

I am deeply indebted to Helena Kole\v sov\' a for finding the error
in the original treatment of Majorana masses of sterile neutrinos
and for suggesting the correct one. The work on this project has
been supported by the grant LG 15052 of the Ministry of Education of
the Czech Republic.

\vspace{4mm}

\section{Appendix: Dynamical fermion mass generation}

In the Appendix we present solutions of the gap equations
(\ref{gammaD}) and (\ref{gammaM}) with fermion mixing neglected
($U=V=1$ for the Dirac fermions and $U=1$ for Majorana neutrinos),
respectively:
\begin{equation}
\gamma=\tfrac{1}{4}\lambda_a g_{ab} I(\gamma)\lambda_b\label{A1}
\end{equation}
\begin{equation}
\gamma=-\tfrac{1}{4}\lambda_a g_{ab}
I(\gamma)\lambda_b^{T}\label{A2}
\end{equation}
Here $\lambda_a$ are the Gell-Mann matrices and $\gamma$ is a real
diagonal matrix with positive entries which determines the fermion
masses as follows. Because $\Sigma(p^2)\equiv
\tfrac{\Lambda^2}{p}\gamma$, the fermion mass, defined as a pole of
the full fermion propagator is
\begin{equation*}
m=\Sigma(p^2=m^2)=\Lambda \gamma^{1/2}
\end{equation*}

With $g_{11}, g_{22}, g_{33}, g_{38}, g_{44}, g_{55}, g_{66},
g_{77}, g_{88}$ different from zero the right hand sides of
equations (\ref{A1}) and (\ref{A2}) are the diagonal matrices. The
equations themselves can be rewritten as
\begin{equation}\label{A3}
\gamma_i^{D/M}=\sum_{k=1}^3 \alpha_{ik}^{D/M}\gamma_k^{D/M}
\ln\frac{1+(\gamma_k^{D/M})^2}{(\gamma_k^{D/M})^2}
\end{equation}
where
\begin{widetext}
\begin{equation}\label{alpha}
\alpha^{D/M}=\frac{3}{64\pi^2}\left(\begin{array}{ccc}
\pm\left(g_{33}+\frac{2}{\sqrt{3}}g_{38}+\frac{1}{3}g_{88}\right) & g_{22}\pm g_{11} & g_{55}\pm g_{44}\\
g_{22}\pm g_{11} & \pm\left(g_{33}-\frac{2}{\sqrt{3}}g_{38}+\frac{1}{3}g_{88}\right) & g_{77}\pm g_{66}\\
g_{55}\pm g_{44} & g_{77}\pm g_{66} & \pm \frac{4}{3}g_{88}
\end{array}
\right)
\end{equation}
\end{widetext}
and the upper and lower signs correspond to the Dirac fermion masses
and the Majorana neutrino masses, respectively.

The goal is an immodest one: Demonstrate convincingly that there is
a reasonable set of the effective low-momentum couplings $g_{ab}$,
which  gives rise to the huge masses of Majorana neutrinos
($M_{iR}\sim O(\Lambda)$) and at the same time to the hierarchical
spectrum of many orders of magnitude lower masses $m_i(f)\ll
\Lambda$ of the electroweakly interacting fermions.

Simplifying as much as we can we consider only
\begin{equation*}
g_{33},g_{38},g_{88};g_{11}=-g_{22},g_{44}=-g_{55},g_{66}=-g_{77}
\end{equation*}
different from zero.

(A) The matrix gap equation for the Dirac masses $m_i$ becomes
diagonal and decoupled, and it is easily solved. Provided the
combinations
\begin{eqnarray*}
\alpha_{11}&=&\tfrac{3}{64\pi^2}(g_{33}+\tfrac{2}{\sqrt
3}g_{38}+\tfrac{1}{3}g_{88})\\
\alpha_{22}&=&\tfrac{3}{64\pi^2}(g_{33}-\tfrac{2}{\sqrt
3}g_{38}+\tfrac{1}{3}g_{88})\\
\alpha_{33}&=&\tfrac{3}{64\pi^2}\tfrac{4}{3}g_{88}
\end{eqnarray*}
are all positive and all $\alpha_{ii} \ll 1$, the resulting Dirac
mass formulas are
\begin{equation}
m_i=\Lambda \phantom{b} \rm exp \phantom{b}
(-1/4\alpha_{ii})\label{mD}
\end{equation}

\vspace{3mm}

(B) Finding the solution for the Majorana masses is less
straightforward. First, for $g_{11}=g_{44}=g_{66}=0$, the gap
equations for the Majorana masses have no solution because of the
minus sign in front of the $\alpha_{ii}$. Consequently, $(g_{11},
g_{44}, g_{66})\neq 0$. Second, in the case of sterile Majorana
neutrinos we are not aware of the necessity of the hierarchical mass
spectrum. With the constants $\alpha_{ii}$ fixed by the numerical
values of the Dirac masses the equations (\ref{A3}) for $\gamma_i^M$
can be viewed as a system of three inhomogeneous linear equations
for the unknown $(g_{11},g_{44},g_{66})$:
\begin{widetext}
$$
-\frac{1}{2}\left(\begin{array}{ccc}
I(\gamma_2^M)&I(\gamma_3^M)&0\\
I(\gamma_1^M)&0& I(\gamma_3^M)\\
0&I(\gamma_1^M)&I(\gamma_2^M)
\end{array}\right)
\left(\begin{array}{c} g_{11} \\g_{44} \\g_{66}
\end{array}\right)=
\left(\begin{array}{c}
\gamma_1^M+\tfrac{16\pi^2}{3}\alpha^{D}_{11} I(\gamma_1^M)\\
\gamma_2^M+\tfrac{16\pi^2}{3}\alpha^{D}_{22} I(\gamma_2^M)\\
\gamma_3^M+\tfrac{16\pi^2}{3}\alpha^{D}_{33} I(\gamma_3^M)
\end{array}\right).
$$
\end{widetext}
This set of equations has a solution for any set of $\gamma_i^M >
0$.

To be explicit, let us put for an illustration
$(\gamma_1^M,\gamma_2^M,\gamma_3^M)=(0.1,0.2,0.3)$ and
$(\gamma_1^D,\gamma_2^D,\gamma_3^D)=(10^{-20},10^{-22},10^{-26})$
(this corresponds approximately to the hierarchy for charged leptons
provided $\Lambda=10^{10}\,$GeV). Then
\begin{widetext}
$$g=\left(
\begin{array}{cccccccc}
 8.08101 & 0 & 0 & 0 & 0 & 0 & 0 & 0 \\
 0 & -8.08101 & 0 & 0 & 0 & 0 & 0 & 0 \\
 0 & 0 & 1.7425 & 0 & 0 & 0 & 0 & 0.0899893 \\
 0 & 0 & 0 & -21.8124 & 0 & 0 & 0 & 0 \\
 0 & 0 & 0 & 0 & 21.8124 & 0 & 0 & 0 \\
 0 & 0 & 0 & 0 & 0 & -34.029 & 0 & 0 \\
 0 & 0 & 0 & 0 & 0 & 0 & 34.029 & 0 \\
 0 & 0 & 0.0899893 & 0 & 0 & 0 & 0 & 1.31887 \\
\end{array}
\right)$$
\end{widetext}
It is important that the precise size and hierarchy of $\gamma_i^D$
does not play any important role for the numerical values of
$\gamma_i^M$.

We are far from making any strong conclusions from the solutions of
the SD equation found here. They are nevetheless suggestive in the
following respect: Often and naturally the observed hierarchy of
fermion mass scales is attributed to 'tumbling' in asymptotically
free gauge theories \cite{tumbling}: When the gauge coupling in the
most attractive channel, growing towards infrared exceeds the
critical value, the fermion-antifermion condensate (fermion mass M)
is dynamically generated. The gauge coupling starts growing again
towards smaller momenta until it reaches the critical value in the
second most attractive channel, and another fermion-antifermion
condensate (fermion mass $m<M$) is generated. It is not excluded
that, as suggested by our simple analysis, the Majorana and Dirac
masses are in fact dynamically generated at different scales.

\end{document}